\documentclass[aps,twocolumn,preprintnumbers,showpacs,showkeys,nofootinbib%
]{revtex4}
\usepackage{epsfig}
\usepackage{amssymb,amsmath,amsfonts,amsthm,graphicx,psfrag}
\newcommand{\noi}{\noindent}
\newcommand{\beq}{\begin{equation}}
\newcommand{\eeq}{\end{equation}}
\newcommand{\bea}{\begin{eqnarray}}
\newcommand{\eea}{\end{eqnarray}}

\newcommand{\Tab}[1]{Table~\ref{#1}}

\newcommand{\tr}{\operatorname{Tr}}

\begin{document}
\preprint{}

\title{Correlations of Abelian monopoles in quark-gluon plasma}

\author{V.~V.~Braguta}
\email[]{braguta@itep.ru}
\affiliation{High Energy Physics Institute, 142280 Protvino, Russia\\
and Institute of Theoretical and Experimental Physics, 117259 Moscow, Russia}

\author{A.~Yu.~Kotov}
\email[]{kotov@itep.ru}
\affiliation{Institute of Theoretical and Experimental Physics, 117259 Moscow, Russia}

\begin{abstract}
In this paper the properties of thermal Abelian monopoles in the deconfinement phase of the
$SU(2)$ gluodynamics are considered. In particular, to study the properties of the Abelian monopole component of QGP we calculate three-point 
correlation functions of monopoles for different temperatures from the region $T/T_c \in (1.5, 6.8)$.  The results 
of the calculation show that the three-point correlation functions can be described by independent pair correlations 
of monopoles. From this one can conclude that the system of Abelian monopoles in QGP reveals the properties 
of a dilute gas. In addition, one can assert that the interaction between Abelian monopoles is a pair interaction and there are
no three-particle forces acting between monopoles.
\end{abstract}



\maketitle

  One of the most interesting results obtained at RHIC is a large elliptic flow \cite{Adams:2005dq, Adcox:2004mh}. 
Interpretation of this result suggests that Quark-Gluon Plasma (QGP) reveals the properties of strongly 
correlated system with very small shear viscosity \cite{Teaney:2009qa}. An interesting 
explanation of this peculiarity can be given within the hypothesis that unusual properties of QGP are closely 
connected with the magnetic degrees of freedom \cite{Liao:2006ry, Chernodub:2006gu, Chernodub:2009rt, Liao:2008jg, Shuryak:2008eq}.

  In paper \cite{Chernodub:2006gu} such magnetic degrees of freedom have been related to thermal Abelian monopoles evaporating from
the magnetic condensate which is believed to induce color confinement at low temperatures. Moreover it
has been proposed to detect such thermal monopoles in finite temperature lattice QCD simulations, by
identifying them with monopole currents having a non-trivial wrapping in the Euclidean temporal 
direction \cite{Chernodub:2006gu, Bornyakov:1991se, Ejiri:1995gd}.

  The way one can study the monopoles' properties on the lattice is via an Abelian projection after fixing 
the maximally Abelian gauge (MAG) \cite{'tHooft:1981ht, 'tHooft:1982ns}. This gauge as well as 
the properties of monopole clusters have been investigated in  numerous papers both at zero and nonzero temperature (see for extensive
list of references, e.g. \cite{Ripka:2003vv}). The evidence was found that the nonperturbative properties of the 
gluodynamics, such as confinement, deconfining transition, chiral symmetry breaking, etc. are closely related
to the Abelian monopoles defined in MAG. This was called a monopole dominance.

  Motivated by the hypothesis that thermal Abelian monopoles might be responsible for the unusual properties of QGP
in this paper we continue the study of their properties. In particular, we are going to study monopole 
correlation functions in order to address the question of collective 
phenomena of magnetic component of QGP. Although the study of two-point correlation 
functions  carried out in papers \cite{D'Alessandro:2007su, Bornyakov:2011th, Bornyakov:2011eq, Bornyakov:2011dj}
revealed rather nontrivial interaction between  monopoles, it is rather difficult to 
draw some conclusion about the properties of monopole medium in QGP. To study
the properties of this medium in this paper we consider three-point correlation 
functions of monopoles.

The correlation function under consideration can be defined as follows
\beq
g^{(3)}(r_{12},r_{13},r_{23})=\left.\frac{\langle\rho(\bar {r}_1)\rho(\bar{r}_2)\rho(\bar{r}_3)\rangle}{\rho^3}\right.,
\label{cor3}
\eeq
where $\bar {r}_1, \bar {r}_2, \bar {r}_3,$ are the positions of three monopoles, 
$r_{12}=|\bar {r}_1 - \bar {r}_2 |,~r_{13}=|\bar {r}_1 - \bar {r}_3 |,~r_{23}=|\bar {r}_2 - \bar {r}_3 | $ are distances 
between the monopoles, 
$\rho(\bar{r})$ is the  operator of monopole density at the point $\bar{r}$ and $\rho$ is the averaged density.

To study collective phenomena and medium effects we are going to compare correlation function (\ref{cor3}) with 
the model correlation function 
\beq
G^{(3)}(r_{12},r_{13},r_{23})=g^{(2)}(r_{12}) g^{(2)}(r_{13}) g^{(2)}(r_{23}),
\label{cor2}
\eeq
where $g^{(2)} (r)$ is the two-point correlation function 
\beq
g^{(2)}(r_{12}) = \frac {\langle \rho(\bar {r}_1) \rho(\bar {r}_2) \rangle} {\rho^2},
\eeq
which will be taken from paper \cite{Bornyakov:2011eq}. Now two comments are in order:
\begin{enumerate}

\item Model (\ref{cor2}) implies
that three-particle correlation takes place only through independent correlation of the pairs. 
Such correlation function is valid for the systems similar to a dilute gas. 
Evidently in a dilute gas there are no collective phenomena and one can 
disregard the influence of monopole medium to the system of three monopoles. 
So, the deviation of correlation function (\ref{cor3}) from model 
function (\ref{cor2}) can be considered as a measure of collective 
phenomena and monopole medium effects. 

\item From correlation function (\ref{cor2}) one can conclude that the 
interaction between monopoles in the monopole medium is a pair interaction 
 described by some universal potential $V(r)$, which can be extracted 
from two-point correlation function. The potential $V(r)$ depends on 
the distance between two monopoles $r$ and the temperature of QGP. 
We believe that last property is rather non-trivial property 
of nonabelian gluodynamics.

\end{enumerate}

To model the system of Abelian monopoles in QGP we use $SU(2)$ lattice gauge theory with 
the standard Wilson action
\beq
S  = \beta \sum_x\sum_{\mu >\nu}
\left[ 1 -\frac{1}{2}\tr \Bigl(U_{x\mu}U_{x+\mu;\nu}
U_{x+\nu;\mu}^{\dagger}U_{x\nu}^{\dagger} \Bigr)\right], \nonumber
\label{eq:action}
\eeq
\noi where $\beta = 4/g_0^2$ and $g_0$ is a bare coupling constant.

\noi Our calculations were performed on the asymmetric lattices with
lattice volume $V=L_t L_s^3$, where $L_{t,s}$ is the number of sites in
the time (space) direction. The temperature $T$ is given by

\beq
T = \frac{1}{aL_t}~,
\eeq

\noi where $a$ is the lattice spacing.

The MAG is fixed by finding an extremum of the gauge functional

\beq
F_U(g) = ~\frac{1}{4V}\sum_{x\mu}~\frac{1}{2}~\tr~\biggl( U^{g}_{x\mu}\sigma_3 U^{g\dagger}_{x\mu}\sigma_3 \biggr) \;,
\label{eq:gaugefunctional}
\eeq

\noi with respect to gauge transformations $g_x$. We apply the simulated annealing (SA) algorithm which proved to be very efficient for this
gauge \cite{Bali:1996dm} as well as for other gauges such as center gauge \cite{Bornyakov:2000ig} and Landau gauge \cite{Bogolubsky:2007bw}.

In \Tab{tab:statistics} we provide the information about the gauge field ensembles used in our study.


\begin{table}[ht]
\begin{center}
\begin{tabular}{|c|c|c|c|c|c|} \hline
 $\beta$ & $a$[fm] & $L_t$ & $~L_s~$ & $T/T_c$ &
$N_{meas}$ \\ \hline\hline

2.43	& 0.108	& 4	& 32	& 1.5	& 1000	\\
2.635	& 0.054	& 4	& 36	& 3.0	& 500	\\
2.80	& 0.034	& 4	& 48	& 4.8	& 1000	\\
2.93	& 0.024	& 4	& 48	& 6.8	& 1000	\\ 
  \hline
\end{tabular}
\end{center}
\caption{Values of $\beta$, lattice sizes, temperatures, number of
configurations. To fix the scale we take $\sqrt{\sigma}=440$ MeV.
}
\label{tab:statistics}
\medskip \noindent
\end{table}


Lattice version of correlation  functions (\ref{cor3}) can be written as follows
\beq
g^{(3)}(r_1,r_2,r_3)=\frac1{\rho^3} \frac{dN(r_1,r_2,r_3)}{dV(r_1,r_2,r_3)}
\label{3point}
\eeq 
where $dN(r_1,r_2,r_3)$ is the total number of triples of monopoles such that the distances between monopoles 
lie in the domain $r_{12} \in (r_1,r_1+\Delta r)$, $r_{13} \in (r_2,r_2+\Delta r)$, $r_{23} \in (r_3,r_3+\Delta r)$. 
 $dV(r_1,r_2,r_3)$ is the number of lattice cubes located in the same domain. 
In order to take into 
account discretization errors we evaluate the $dV(r_1,r_2,r_3)$ numerically. $\Delta r$ is the size of one bin.
Additional factor  $1/\rho^3$  was introduced  to normalize the whole expression. 
At large distances, where there are no correlations at all, $g=1$. 

\begin{figure}[t]
\centering
\includegraphics[width=6.0cm,angle=270]{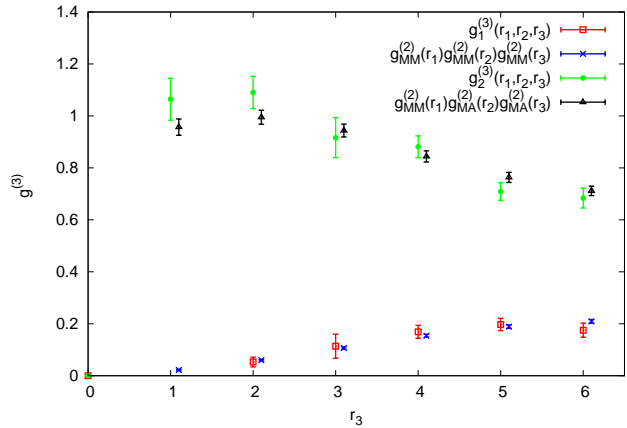}
\caption{$\beta=2.43, T/T_c=1.5$. The correlation function $g^{(3)}(r_1,r_2,r_3)$ and the model function $G^{(3)}(r_1,r_2,r_3)$. 
The distances between the first and second, the first and the third monopoles are $r_1=r_2=3$ lattice spacings. 
The distance between the second and the third monopoles $r_3$ is varied. 
}
\label{fig:total_r3_b2.43}
\end{figure}

In our analysis only monopoles with magnetic charge $q = \pm 1$ are taken into account. Our results show 
that the monopoles with $|q|>1$ are greatly suppressed. Since one considers only two types of effective particles 
(monopoles $q=+1$ and antimonopoles $q=-1$) there are four different correlators $g^{(3)}_{MMM}, g^{(3)}_{AAA}, g^{(3)}_{MMA}, g^{(3)}_{AAM}$, where 
$M, A$ denote monopole and antimonopole. Evidently, monopoles are equivalent to antimonopoles in the sense that one 
can make magnetic charge conjugation and this does not change the physical properties 
of QGP. For this reason, instead of four correlators we calculate the following linear combinations
\beq
g^{(3)}_1 = \frac 1 2 \bigl (g^{(3)}_{MMM} + g^{(3)}_{AAA} \bigr ),~~~g^{(3)}_2 = \frac 1 2 \bigl (g^{(3)}_{MMA} + g^{(3)}_{AAM} \bigr )
\label{cor_last}
\eeq

\begin{figure}[t]
\centering
\includegraphics[width=6.0cm,angle=270]{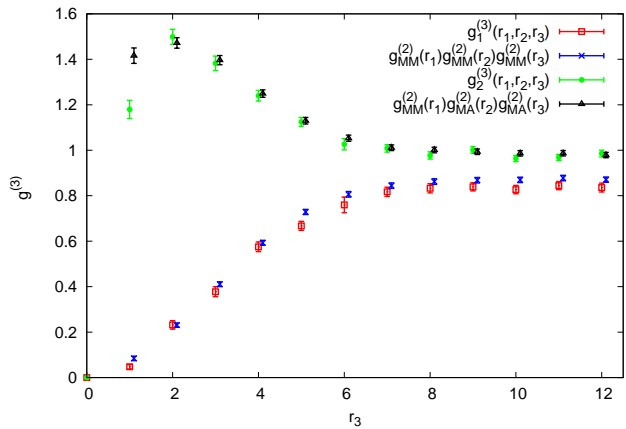}
\caption{$\beta=2.43, T/T_c=1.5$. The correlation function $g^{(3)}(r_1,r_2,r_3)$ and the model function $G^{(3)}(r_1,r_2,r_3)$.
The distances between the first and second, the first and the third monopoles are $r_1=r_2=6$ lattice spacings. 
The distance between the second and the third monopoles $r_3$ is varied. 
}
\label{fig:total_r6_b2.43}
\end{figure}

\begin{figure}[tb]
\centering
\includegraphics[width=6.0cm,angle=270]{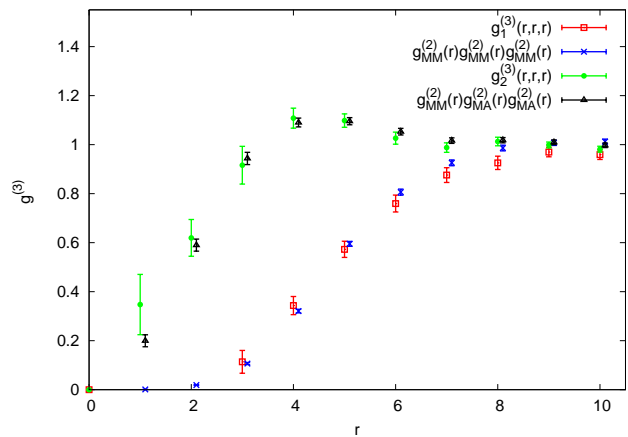}
\caption{$\beta=2.43, T/T_c=1.5$. The correlation function $g^{(3)}(r_1,r_2,r_3)$ and model function $G^{(3)}(r_1,r_2,r_3)$. 
 The monopoles are located at the corners of the regular triangle 
$r_1=r_2=r_3=r$ and the side of this triangle $r$ is varied.
}
\label{fig:triangle_b2.43}
\end{figure}

Now let us proceed to the results of this paper. In Figures~\ref{fig:total_r3_b2.43}, \ref{fig:total_r6_b2.43} and \ref{fig:triangle_b2.43} 
we plot the correlation functions $g_1^{(3)}(r_1,r_2,r_3), g_2^{(3)}(r_1,r_2,r_3)$ and model (\ref{cor2}) for the configurations with 
$\beta=2.43, T/T_c=1.5$. In Fig.~\ref{fig:total_r3_b2.43} we fixed the distances between the first and second, the first and
the third monopoles at the values $r_1=r_2=3$ lattice spacings and varied the distance between the second and the third monopoles $r_3$. 
Similarly, in Fig. \ref{fig:total_r6_b2.43} we take the $r_1=r_2=6$ lattice spacings and varied the $r_3$. 
In Fig. \ref{fig:triangle_b2.43} the monopoles are located at the corners of a regular triangle 
$r_1=r_2=r_3=r$ and the side of this triangle $r$ is varied. From these figures it is seen that up to the statistical uncertainty 
the correlation functions $g_1, g_2$  coincide with the corresponding models (\ref{cor2}).
Similar conclusion can be drawn for the other temperatures $T/T_c=3.0, 4.8, 6.8$  studied in this paper.

Figures~\ref{fig:total_r3_b2.43}, \ref{fig:total_r6_b2.43} and \ref{fig:triangle_b2.43} can give us only qualitative 
result. To get the quantitative measurement of the discrepancy between models (\ref{cor3}) and (\ref{cor2}) we 
introduce the following quantity
\beq
\delta=\frac1{N}\sum_{r_1,r_2,r_3}\frac{(g^{(3)}(r_1,r_2,r_3)-G^{(3)}(r_1,r_2,r_3))^2}{\sigma^2(r_1,r_2,r_3)}
\label{chi_squared}
\eeq 
Here $\sigma(r_1,r_2,r_3)$ is the uncertainty of the calculation of the correlation function $g^{(3)}(r_1,r_2,r_3)$ at the given point $(r_1,r_2,r_3)$. 
Note that we have disregarded the uncertainty in the two-point functions since it is small as compared to 
the $\sigma(r_1,r_2,r_3)$.

There are some restrictions on the values of  distances between monopoles. The first one comes from the finite volume effect. 
Evidently, if $r_1+r_2+r_3>L_s$ then due to periodical boundary conditions new non-physical triples of monopoles wrapped in a spatial 
direction appear. We ignored such configurations in the calculation. 
Another restriction comes from the triangle inequality: $|r_1-r_2|<r_3<r_1+r_2$. We also did not take into account 
the distances smaller than $3$ lattice spacings  due to the large statistical uncertainty. In formula (\ref{chi_squared}) the sum is taken over all 
distances with the mentioned restrictions. $N$ is the total number of triples of distances that satisfy these restrictions. 
Obviously the value of the $\delta$ is $\approx1$ if there is no discrepancy between two correlation functions. 
In \Tab{tab:chi2} we present the values of the $\delta$ for the different $\beta$. From this Table it is seen that 
up to the uncertainty of the calculation the three-point correlation function of Abelian monopoles can be 
described by model (\ref{cor2}).

\begin{table}[ht]
\begin{center}
\begin{tabular}{|c|c|c|c|c|c|} \hline
 $\beta$ & $T/T_c$ & $\delta_1$ & $\delta_2$ \\ \hline\hline
  2.43    & 1.5 & 0.59 & 0.67 \\
  2.635   & 3.0 & 0.67 & 0.68 \\
  2.80   & 4.8 & 0.66 & 0.69 \\
  2.93    & 6.8 & 0.62 & 0.68 \\\hline
\end{tabular}
\end{center}
\caption{The values of $\beta$, temperature, the difference $\delta$ between correlation functions (\ref{cor3}) and the 
corresponding models (\ref{cor2}). 
The $\delta_1$ corresponds to the correlation function $g_1$ in (\ref{cor_last}), 
The $\delta_2$ corresponds to the correlation function $g_2$ in (\ref{cor_last}).
}
\label{tab:chi2}
\medskip \noindent
\end{table}


{\it In conclusion}, in this paper we studied the properties of thermal Abelian monopoles in the deconfinement phase of the
$SU(2)$ gluodynamics. In particular, to study the properties of Abelian monopole component in QGP we calculated three-point 
correlation functions of monopoles for different temperatures from the region $T/T_c \in (1.5, 6.8)$.  The results 
of the calculation show that the tree-point correlation functions can be described by the independent pair correlation 
of monopoles. From the last fact one can conclude that { \it the system of Abelian monopoles in QGP reveals the properties 
of dilute gas.} In addition, one can assert that the interaction between Abelian monopoles is a pair interaction and there are 
no three-particle forces acting between monopoles.
\subsection*{Acknowledgments}
We would like to express our gratitude to M.I. Polikarpov, V.G. Bornyakov and V.I. Zakharov for very useful discussions.
This investigation has been supported by the Federal Special-Purpose
Program 'Cadres' of the Russian Ministry of Science and Education (contracts 02.740.11.0571, No. 02.740.11.0490) and by grant
RFBR 11-02-01227-a. VVB is supported by grant RFBR 10-02-00061-a and RFBR 11-02-00015-a.


\begin{thebibliography}{21}
\expandafter\ifx\csname natexlab\endcsname\relax\def\natexlab#1{#1}\fi
\expandafter\ifx\csname bibnamefont\endcsname\relax
  \def\bibnamefont#1{#1}\fi
\expandafter\ifx\csname bibfnamefont\endcsname\relax
  \def\bibfnamefont#1{#1}\fi
\expandafter\ifx\csname citenamefont\endcsname\relax
  \def\citenamefont#1{#1}\fi
\expandafter\ifx\csname url\endcsname\relax
  \def\url#1{\texttt{#1}}\fi
\expandafter\ifx\csname urlprefix\endcsname\relax\def\urlprefix{URL }\fi
\providecommand{\bibinfo}[2]{#2}
\providecommand{\eprint}[2][]{\url{#2}}


\bibitem[\citenamefont{Adams et~al.}(2005)]{Adams:2005dq}
\bibinfo{author}{\bibfnamefont{J.}~\bibnamefont{Adams} {\it et al.} [STAR Collaboration]}
  \bibinfo{journal}{Nucl. Phys. A} \textbf{\bibinfo{volume}{757}},
\bibinfo{pages}{102} (\bibinfo{year}{2005}),
  \eprint{nucl-ex/0501009}.

\bibitem[\citenamefont{Adcox et~al.}(2004)]{Adcox:2004mh}
\bibinfo{author}{\bibfnamefont{K.}~\bibnamefont{Adcox}  {\it et al.} [PHENIX Collaboration]},
  \bibinfo{journal}{Nucl. Phys. A} \textbf{\bibinfo{volume}{757}},
\bibinfo{pages}{184} (\bibinfo{year}{2005}),
  \eprint{nucl-ex/0410003}.

\bibitem[\citenamefont{Teaney}(2009)]{Teaney:2009qa}
\bibinfo{author}{\bibfnamefont{D.~A.}~\bibnamefont{Teaney}},
\eprint{nucl-th/0905.2433}.

\bibitem[\citenamefont{Chernodub et~al.}(2009)\citenamefont{Chernodub, Verschelde and Zakharov}]{Chernodub:2009rt}
\bibinfo{author}{\bibfnamefont{M.~N.}~\bibnamefont{Chernodub}},
  \bibinfo{author}{\bibfnamefont{H.}~\bibnamefont{Verschelde}}, \bibnamefont{and}
  \bibinfo{author}{\bibfnamefont{V.~I.}~\bibnamefont{Zakharov}}  
\bibinfo{journal}{Nucl. Phys. Proc. Suppl.} \textbf{\bibinfo{volume}{207-208}},
\bibinfo{pages}{325} (\bibinfo{year}{2010}),
\eprint{hep-ph/0905.2520}
 
\bibitem[\citenamefont{Liao and Shuryak}(2008)]{Liao:2008jg}
\bibinfo{author}{\bibfnamefont{J.}~\bibnamefont{Liao}} \bibnamefont{and}
\bibinfo{author}{\bibfnamefont{E.}~\bibnamefont{Shuryak}},
\bibinfo{journal}{Phys. Rev. Lett.} \textbf{\bibinfo{volume}{101}},
\bibinfo{pages}{162302} (\bibinfo{year}{2008}),
\eprint{hep-ph/0804.0255}

\bibitem[{\citenamefont{Liao and Shuryak}(2007)}]{Liao:2006ry}
\bibinfo{author}{\bibfnamefont{J.}~\bibnamefont{Liao}} \bibnamefont{and}
  \bibinfo{author}{\bibfnamefont{E.}~\bibnamefont{Shuryak}},
  \bibinfo{journal}{Phys. Rev.} \textbf{\bibinfo{volume}{C75}},
  \bibinfo{pages}{054907} (\bibinfo{year}{2007}), \eprint{hep-ph/0611131}.
  
  \bibitem[{\citenamefont{Chernodub and Zakharov}(2007)}]{Chernodub:2006gu}
\bibinfo{author}{\bibfnamefont{M.~N.} \bibnamefont{Chernodub}}
  \bibnamefont{and} \bibinfo{author}{\bibfnamefont{V.~I.}
  \bibnamefont{Zakharov}}, \bibinfo{journal}{Phys. Rev. Lett.}
  \textbf{\bibinfo{volume}{98}}, \bibinfo{pages}{082002}
  (\bibinfo{year}{2007}), \eprint{hep-ph/0611228}.

\bibitem[{\citenamefont{Shuryak}(2009)}]{Shuryak:2008eq}
\bibinfo{author}{\bibfnamefont{E.}~\bibnamefont{Shuryak}},
  \bibinfo{journal}{Prog. Part. Nucl. Phys.} \textbf{\bibinfo{volume}{62}},
  \bibinfo{pages}{48} (\bibinfo{year}{2009}), \eprint{0807.3033}.
   
\bibitem[\citenamefont{Bornyakov}(1991)]{Bornyakov:1991se}
\bibinfo{author}{\bibfnamefont{V.~G.}~\bibnamefont{Bornyakov}},
\bibinfo{author}{\bibfnamefont{V.~K.}~\bibnamefont{Mitrjushkin}} \bibnamefont{and}
\bibinfo{author}{\bibfnamefont{M.}~\bibnamefont{Muller-Preussker}}
\bibinfo{journal}{Phys. Lett. B} \textbf{\bibinfo{volume}{284}},
\bibinfo{pages}{99} (\bibinfo{year}{1992}).  

\bibitem[\citenamefont{Ejiri}(1995)]{Ejiri:1995gd}
\bibinfo{author}{\bibfnamefont{S.}~\bibnamefont{Ejiri}},
\bibinfo{journal}{Phys. Lett. B} \textbf{\bibinfo{volume}{376}},
\bibinfo{pages}{163} (\bibinfo{year}{1996}),
\eprint{hep-lat/9510027}.

\bibitem[\citenamefont{'tHooft}(1982)]{'tHooft:1981ht}
\bibinfo{author}{\bibfnamefont{G.}~\bibnamefont{'t Hooft}},
\bibinfo{journal}{Nucl. Phys. B} \textbf{\bibinfo{volume}{190}},
\bibinfo{pages}{455} (\bibinfo{year}{1981}).


\bibitem[\citenamefont{'tHooft}(1982)]{'tHooft:1982ns}
\bibinfo{author}{\bibfnamefont{G.}~\bibnamefont{'t Hooft}},
  \bibinfo{journal}{Phys. Scripta} \textbf{\bibinfo{volume}{25}},
  \bibinfo{pages}{133} (\bibinfo{year}{1982}).

\bibitem[\citenamefont{Ripka}(2003)]{Ripka:2003vv}
\bibinfo{author}{\bibfnamefont{G.}~\bibnamefont{Ripka}},
\eprint{hep-ph/0310102}.

\bibitem[{\citenamefont{D'Alessandro and D'Elia}(2008)}]{D'Alessandro:2007su}
\bibinfo{author}{\bibfnamefont{A.}~\bibnamefont{D'Alessandro}}
  \bibnamefont{and} \bibinfo{author}{\bibfnamefont{M.}~\bibnamefont{D'Elia}},
  \bibinfo{journal}{Nucl. Phys.} \textbf{\bibinfo{volume}{B799}},
  \bibinfo{pages}{241} (\bibinfo{year}{2008}), \eprint{0711.1266}.

\bibitem[\citenamefont{Bornyakov and Kononenko}(2011)]{Bornyakov:2011dj}
\bibinfo{author}{\bibfnamefont{V.~G.}~\bibnamefont{Bornyakov}}
\bibnamefont{and}
\bibinfo{author}{\bibfnamefont{A.~G.}~\bibnamefont{Kononenko}},
\eprint{hep-lat/1111.0169}.

\bibitem[{\citenamefont{Bornyakov and Braguta}(2011a)}]{Bornyakov:2011th}
\bibinfo{author}{\bibfnamefont{V.G.}~\bibnamefont{
Bornyakov}} \bibnamefont{and}
  \bibinfo{author}{\bibfnamefont{V.V.}~\bibnamefont{Braguta}},
  \bibinfo{journal}{Phys. Rev.} \textbf{\bibinfo{volume}{D84}},
  \bibinfo{pages}{074502} (\bibinfo{year}{2011}), \eprint{hep-lat/1104.1063}.
  
  \bibitem[{\citenamefont{Bornyakov and Braguta}(2012)}]{Bornyakov:2011eq}
\bibinfo{author}{\bibfnamefont{V.G.}~\bibnamefont{
Bornyakov}} \bibnamefont{and}
  \bibinfo{author}{\bibfnamefont{V.V.}~\bibnamefont{Braguta}},
  \bibinfo{journal}{Phys. Rev.} \textbf{\bibinfo{volume}{D85}},
  \bibinfo{pages}{014502} (\bibinfo{year}{2012}), \eprint{hep-lat/1110.6308}.

\bibitem[{\citenamefont{Bali et~al.}(1996)\citenamefont{Bali, Bornyakov,
  Muller-Preussker, and Schilling}}]{Bali:1996dm}
\bibinfo{author}{\bibfnamefont{G.}~\bibnamefont{Bali}},
  \bibinfo{author}{\bibfnamefont{V.}~\bibnamefont{Bornyakov}},
  \bibinfo{author}{\bibfnamefont{M.}~\bibnamefont{Muller-Preussker}},
  \bibnamefont{and}
  \bibinfo{author}{\bibfnamefont{K.}~\bibnamefont{Schilling}},
  \bibinfo{journal}{Phys.Rev.} \textbf{\bibinfo{volume}{D54}},
  \bibinfo{pages}{2863} (\bibinfo{year}{1996}), \eprint{hep-lat/9603012}.

\bibitem[{\citenamefont{Bornyakov et~al.}(2001)\citenamefont{Bornyakov,
  Komarov, and Polikarpov}}]{Bornyakov:2000ig}
\bibinfo{author}{\bibfnamefont{V.}~\bibnamefont{Bornyakov}},
  \bibinfo{author}{\bibfnamefont{D.}~\bibnamefont{Komarov}}, \bibnamefont{and}
  \bibinfo{author}{\bibfnamefont{M.}~\bibnamefont{Polikarpov}},
  \bibinfo{journal}{Phys. Lett.} \textbf{\bibinfo{volume}{B497}},
  \bibinfo{pages}{151} (\bibinfo{year}{2001}), \eprint{hep-lat/0009035}.

\bibitem[{\citenamefont{Bogolubsky et~al.}(2008)\citenamefont{Bogolubsky,
  Bornyakov, Burgio, Ilgenfritz, Muller-Preussker et~al.}}]{Bogolubsky:2007bw}
\bibinfo{author}{\bibfnamefont{I.}~\bibnamefont{Bogolubsky}},
  \bibinfo{author}{\bibfnamefont{V.}~\bibnamefont{Bornyakov}},
  \bibinfo{author}{\bibfnamefont{G.}~\bibnamefont{Burgio}},
  \bibinfo{author}{\bibfnamefont{E.}~\bibnamefont{Ilgenfritz}},
  \bibinfo{author}{\bibfnamefont{M.}~\bibnamefont{Muller-Preussker}},
  \bibnamefont{et~al.}, \bibinfo{journal}{Phys.Rev.}
  \textbf{\bibinfo{volume}{D77}}, \bibinfo{pages}{014504}
  (\bibinfo{year}{2008}), \eprint{0707.3611}.
  
\end{thebibliography}
\end{document}